# From Digital Capabilities to Non-Linear Growth: The Exponential SME Framework

**Dhushy Thillaivasan**

School of Business, Melbourne Institute of Technology, Melbourne, Australia

Email: DThillaivasa@mit.edu.au

*Corresponding author

**Kristina Nicholls**

School of Business, Melbourne Institute of Technology, Melbourne, Australia

Email: knicholls@mit.edu.au

**Sathyapriya Govindarajulu**

School of Business, Melbourne Institute of Technology, Melbourne, Australia

Email: SGovindarajulu@mit.edu.au

**Ken Mardaneh**

School of Business, Melbourne Institute of Technology, Melbourne, Australia

Email: kmardaneh@mit.edu.au

## Abstract

Small and medium-sized enterprises (SMEs) face increasing pressure to navigate digital transformation, yet many struggle to convert digital initiatives into sustainable organisational capability development and growth outcomes. This paper develops the *Exponential SME Framework*, an integrated conceptual framework that explains how SMEs may progress from digital capability development to non-linear growth potential. Drawing on literature relating to Digital Dynamic Capabilities (DDC), digital transformation practices, organisational transformation, Exponential Organization (ExO) attributes, and SME growth, the framework synthesises previously fragmented streams of research into a coherent conceptual structure. The framework proposes that Digital Dynamic Capabilities provide the foundation for Digital Transformation Practices (DASAT), which contribute to transformations across key organisational domains, including business models, value chains, customer relationships, and organisational culture. These transformations may support the adoption of ExO-like attributes, specifically experimentation, dashboards, autonomy, and community engagement, that enhance organisational adaptability, learning, and responsiveness. Through the interaction of these organisational elements, SMEs may strengthen the conditions associated with non-linear growth potential. The paper contributes to the SME and digital transformation literature by integrating multiple streams of research into a unified framework, clarifying the relationships between capabilities, practices, organisational transformation, and growth-enabling mechanisms, and providing a foundation for future empirical research. The paper also offers managerial and policy implications for strengthening SME digital transformation and growth readiness.

## 1. Introduction

Small and medium-sized enterprises (SMEs) are widely recognised as major contributors to economic development, innovation, employment creation, and regional competitiveness. Despite their significance, many SMEs struggle to sustain growth in increasingly dynamic and digitally enabled business environments. Rapid technological change, evolving customer expectations, intensifying competition, and growing digital disruption have increased pressure on SMEs to adapt their operations, business models, and value creation mechanisms to remain competitive (Ciasullo et al., 2025; Karadağ et al., 2024; Verhoef et al., 2021). Consequently, understanding how SMEs leverage digital transformation to improve organisational performance and unlock new growth opportunities has become an important research priority.

Digital transformation has emerged as a strategic response to these challenges. By integrating digital technologies into organisational processes, products, services, and business models, firms can improve efficiency, enhance customer experiences, increase adaptability, and create new sources of value (Verhoef et al., 2021; Vial, 2019). However, unlike large organisations that often possess dedicated transformation teams, extensive digital expertise, and substantial financial resources, SMEs typically pursue digital transformation under conditions of resource scarcity, limited technical expertise, and informal organisational structures (Li et al., 2018; Matt et al., 2015). As a result, digital transformation within SMEs frequently unfolds through experimentation, learning-by-doing, and incremental organisational adaptation rather than through coordinated enterprise-wide transformation programs.

A substantial body of literature has sought to explain how organisations develop the capacity to navigate digital change. Building on Dynamic Capability Theory (D. J. Teece, 2007) , Digital Dynamic Capabilities (DDC) have been identified as a critical mechanism enabling firms to sense digital opportunities, seize emerging technologies, and reconfigure organisational resources in response to changing market conditions (Ciasullo et al., 2025; Warner & Wäger, 2019). Within SMEs, DDC provide an important foundation for digital transformation by enabling organisations to continuously adapt and respond to technological and environmental change despite resource constraints (Karadağ et al., 2024) . However, while DDC explain how firms develop the capacity for transformation, they provide limited insight into how transformation is enacted and how such efforts contribute to growth-oriented outcomes.

Research on digital transformation practices offers additional insight into how transformation occurs within organisations. Studies have identified a range of practices associated with successful digital transformation, including digital awareness, strategic alignment, technology adoption, and continuous transformation activities (Chanias et al., 2019; Kahveci, 2025). Collectively conceptualised in this paper as DASAT (Kahveci, 2025), these practices represent the operational mechanisms through which digital transformation is implemented. Yet existing research provides limited understanding of how such practices interact with organisational capabilities and contribute to broader organisational renewal.

Beyond capabilities and practices, digital transformation frequently requires changes across multiple organisational domains, including business models, value chains, customer relationships, and organisational culture (Gupta, 2018; Verhoef et al., 2021). Such domain-level transformations are particularly important for SMEs because they influence how value is created, delivered, and captured. However, research examining domain transformation has generally remained disconnected from studies of capability development and transformation practices, making it difficult to understand how these elements collectively contribute to organisational growth.

A further stream of research has examined the characteristics of highly adaptive and scalable organisations. The Exponential Organization (ExO) literature highlights organisational attributes such as experimentation, decentralised autonomy, real-time information visibility, and ecosystem engagement as important enablers of agility, responsiveness, and innovation (Ismail et al., 2014, 2023). Although most SMEs are unlikely to become fully exponential organisations, they may adopt ExO-like attributes that strengthen their ability to respond to change and exploit opportunities (Bresciani et al., 2021). Nevertheless, little is known about how SMEs develop the organisational foundations necessary to adopt these attributes or how such attributes contribute to growth outcomes.

Finally, while growth remains a central objective for many SMEs, existing research suggests that growth rarely occurs in a linear and predictable manner. SMEs often experience periods of incremental progress, accelerated expansion, strategic renewal, and temporary stagnation. Despite increasing interest in digital transformation, relatively little research has examined how digital capabilities, transformation practices, organisational change, and ExO-like attributes collectively contribute to the development of non-linear growth potential within SMEs.

Taken together, these observations reveal a significant gap in the literature. Existing research provides valuable insights into digital capabilities, digital transformation practices, organisational change, ExO-like attributes, and SME growth. However, these streams have largely evolved independently, limiting our understanding of how they interact to support transformative and potentially non-linear growth outcomes. As a result, there remains a need for an integrated framework capable of explaining how SMEs progress from digital capability development to growth-enhancing organisational transformation.

To address this gap, this paper develops the *Exponential SME Framework*, an integrated conceptual framework that synthesises literature on Digital Dynamic Capabilities, digital transformation practices, organisational domain transformation, ExO-like attributes, and SME growth. The framework provides a holistic explanation of how SMEs can leverage digital transformation to strengthen organisational adaptability, develop growth-enabling capabilities, and enhance their potential for non-linear growth.

Accordingly, this paper addresses the following research question: How do SMEs progress from digital capability development to non-linear growth through digital transformation?

This paper contributes to the SME and digital transformation literature in three ways. First, it synthesises fragmented streams of literature relating to Digital Dynamic Capabilities, digital transformation practices, organisational domain transformation, ExO-like attributes, and SME growth into an integrated conceptual framework termed the Exponential SME Framework. Second, it explains how these elements collectively contribute to the development of non-linear growth potential within SMEs. Third, it establishes a future research agenda by identifying important organisational mechanisms and relationships that can be examined, refined, and extended through subsequent empirical investigation.

The remainder of the paper is structured as follows. The next section reviews and synthesises the literature underpinning the Exponential SME Framework. The framework is then developed and its core components are explained. This is followed by a discussion of the theoretical and practical implications of the framework, before concluding with a future research agenda for advancing SME digital transformation and growth research.

## 2. Theoretical Background

To develop an integrated understanding of how SMEs progress from digital capability development to non-linear growth potential, it is necessary to examine the theoretical foundations that underpin each component of this transformation pathway. Existing research spans multiple domains, including dynamic capabilities, digital transformation practices, organisational change, ExO-like mechanisms, and SME growth, but these streams have rarely been synthesised into a coherent, integrated conceptual framework. A structured review of these literatures provides the conceptual grounding needed to explain how capabilities translate into practices, how practices reshape organisational domains, and how these transformed domains enable the adoption of mechanisms that support non-linear growth.

Accordingly, the following section reviews the key theoretical constructs that inform the proposed model: Digital Dynamic Capabilities (DDC), Digital Transformation Practices (DASAT), Organisational Domain Transformation, ExO-like Attribute Adoption, and Non-linear Growth Potential. Together, these constructs form the basis for understanding the conceptual relationships underpinning the Exponential SME Framework.

### 2.1 Digital Dynamic Capabilities (DDC)

Digital Dynamic Capabilities (DDC) refer to a firm's ability to sense digital opportunities and threats, seize digital initiatives, and reconfigure organisational resources and processes in response to technological change (D. J. Teece, 2007; Warner & Wäger, 2019). Rooted in the broader dynamic capability's framework, DDC represent the specific manifestation of these capabilities in digitally intensive environments, where rapid technological shifts require continuous adaptation (D. Teece et al., 2016). For SMEs, DDC are particularly critical because they compensate for structural disadvantages such as limited resources, smaller talent pools, and lower absorptive capacity (Ciasullo et al., 2025; Karadağ et al., 2024).

The sensing dimension of DDC involves identifying digital trends, customer needs, and emerging technologies that may influence the firm's competitive position (Abdurrahman et al., 2024; Kahveci, 2025; Nambisan et al., 2019; D. J. Teece, 2007). The seizing dimension reflects the ability to mobilise resources, make strategic decisions, and invest in digital initiatives that create value (Eller et al., 2020; Kahveci, 2025; Matarazzo et al., 2021; Warner & Wäger, 2019). The transforming dimension encompasses the reconfiguration of processes, structures, and routines to embed digital technologies and support ongoing renewal (Kahveci, 2025; Karadağ et al., 2024; Teece et al., 2016; Verhoef et al., 2021). Together, these capabilities enable SMEs to navigate uncertainty, respond to digital disruption, and pursue new opportunities.

Recent studies emphasise that DDC are not merely technical or IT-related; they are organisational capabilities that integrate leadership, culture, learning, and strategic alignment (Matarazzo et al., 2021; Shin et al., 2023). For SMEs, successful digital transformation is predicated not on the volume of technology acquired, but on the firm's ability to orchestrate change across its internal structures(Rahman et al., 2025; Warner & Wäger, 2019). DDC therefore serve as the foundational enabler of digital transformation practices, shaping how SMEs recognise digital opportunities, prioritise initiatives, and implement change.

In this paper, DDC are conceptualised as the starting point of the transformation pathway. They provide the underlying capacity that enables SMEs to enact digital transformation practices (DASAT), which in turn drive deeper organisational change. Without sufficient DDC, SMEs struggle to move beyond ad hoc digital adoption toward more strategic and integrated transformation efforts.

### 2.2 Digital Transformation Practices (DASAT)

Digital transformation is increasingly understood not as a single event or technology adoption effort, but as an ongoing organisational process enacted through a set of interrelated practices (Chanias et al., 2019; Kahveci, 2025; Vial, 2019). For SMEs, these practices represent the concrete, observable actions through which Digital Dynamic Capabilities (DDC) are expressed and operationalised. While the literature uses varied terminology, such as "digital transformation activities," "digitalisation routines," or "digital strategy enactment", a consistent theme is that transformation unfolds through iterative cycles of awareness, strategising, adoption, and continuous improvement (Chanias et al., 2019; Eller et al., 2020; Kahveci, 2025; Verhoef et al., 2021; Warner & Wäger, 2019).

These practices are conceptualised collectively as DASAT: Digital Awareness, Digital Strategy, Digital Adoption, and Continuous Improvement. This framing reflects the practical, capability-enacting nature of SME digital transformation (Kahveci, 2025)**.** Figure #1 provides a visual representation of the DASAT cycle, illustrating how these four practices unfold as an iterative set of transformation routines within SMEs. While these practices are discussed separately across the digital transformation literature, they are conceptualised collectively in this paper as DASAT (Digital Awareness, Digital Strategy, Digital Adoption, and Continuous Improvement). DASAT is proposed as a synthesising construct that captures the principal organisational practices through which digital transformation is enacted within SMEs (Chanias et al., 2019; Kahveci, 2025; Verhoef et al., 2021; Vial, 2019).

**Figure #1:**

*The DASAT Cycle*

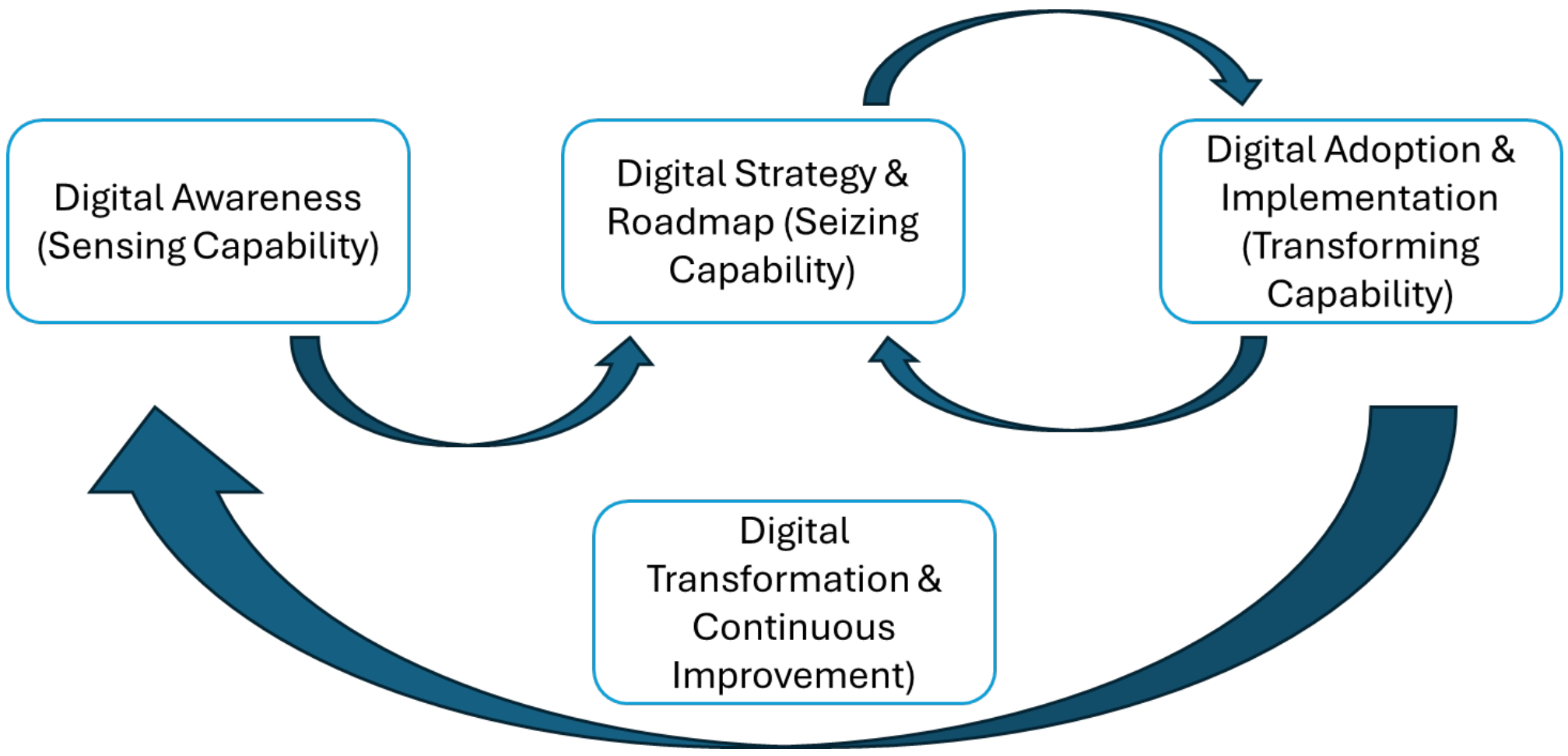

*Note.* Adapted from "Digital Transformation in SMEs: Enablers, Interconnections, and a Framework for Sustainable Competitive Advantage" (Kahveci, 2025). Reproduced and adapted for conceptual illustration purposes.

Digital Awareness refers to the organisation's ability to recognise digital opportunities, threats, and trends relevant to its context (Chanias et al., 2019; Kahveci, 2025). This involves systematic environmental scanning, competitor benchmarking, and an evolving understanding of digitally influenced customer expectations. Such awareness is critical for SMEs, which frequently bypass formalized strategic planning in Favor of managerial intuition and informal sensing mechanisms to navigate digital disruption (Eller et al., 2020; Karadağ et al., 2024; Li et al., 2018).

Digital Strategy involves articulating how digital technologies support organisational goals, value creation, and competitive positioning (Matt et al., 2015). In the SME context, strategy often manifests as an emergent rather than formalized process, heavily influenced by resource scarcity, leadership vision, and opportunistic decision-making (Karadağ et al., 2024; Nambisan et al., 2019). Despite its informal nature, maintaining even a fluid strategic direction is vital; it provides the "lightweight" framework necessary for SMEs to prioritize initiatives, align limited resources, and facilitate ongoing organisational renewal (Abdurrahman et al., 2024; Verhoef et al., 2021).

Digital Adoption encompasses the systemic implementation of digital tools, platforms, and processes across the organization, ranging from cloud-based infrastructures and data analytics to automated, customer-facing workflows(Verhoef et al., 2021). In the SME context, adoption is frequently incremental and trial-based, dictated by the need to balance immediate operational needs with long-term strategic goals (Eller et al., 2020; Karadağ et al., 2024). Despite its gradual nature, the cumulative effect of these implementations significantly enhances operational efficiency and customer engagement, effectively serving as the "Seizing" mechanism of the firm's dynamic capabilities (Matarazzo et al., 2021; Rahman et al., 2025).

Continuous Improvement reflects the iterative refinement of digital processes, routines, and capabilities, reinforcing the principle that digital transformation is an ongoing journey rather than a finite destination (Verhoef et al., 2021; Warner & Wäger, 2019). For SMEs, this evolution is sustained through persistent cycles of experimentation, real-time learning, and organisational adaptation (Kahveci, 2025). Such iterative refinement is essential for maintaining transformation momentum and successfully embedding digital practices into the fabric of daily organisational routines, ensuring long-term resilience (Ciasullo et al., 2025; Karadağ et al., 2024).

Taken together, DASAT represents the practical expression of DDC within SMEs. Whereas DDC provide the underlying capacity to sense, seize, and transform, DASAT captures the actions through which these capabilities are enacted. This distinction is important: SMEs may possess latent digital capabilities, but without the corresponding practices, transformation remains superficial or fragmented.

In the context of this paper, DASAT is conceptualised as the mechanism through which DDC influence organisational change. These practices serve as the bridge between capability development and deeper domain transformation, shaping how SMEs restructure their business models, value chains, customer relationships, and organisational culture.

### 2.3 Organisational Domain Transformation

Digital transformation extends beyond the mere adoption of new technologies; it necessitates fundamental shifts across multiple organisational domains that redefine how firms create, deliver, and capture value (Verhoef et al., 2021). For SMEs, these shifts are particularly consequential, as their internal structures and cultures are often less formalised yet more path-dependent than those of larger incumbents(Eller et al., 2020; Karadağ et al., 2024). Organisational domain transformation thus entails the systemic reconfiguration of the firm's business model, value chain, customer engagement, and organisational culture in response to digital pressures (Gupta, 2018). These domains constitute the core architecture of the firm; the depth of change within them determines whether digital efforts yield only incremental gains or achieve substantive strategic renewal.

Figure #2 illustrates Gupta's (2018) four organisational domains, business model, value chain, customer engagement, and organisational culture, that collectively define the architecture of digital strategy and underpin organisational domain transformation.

**Figure #2:**

*The Four Organisational Domains of Digital Transformation*

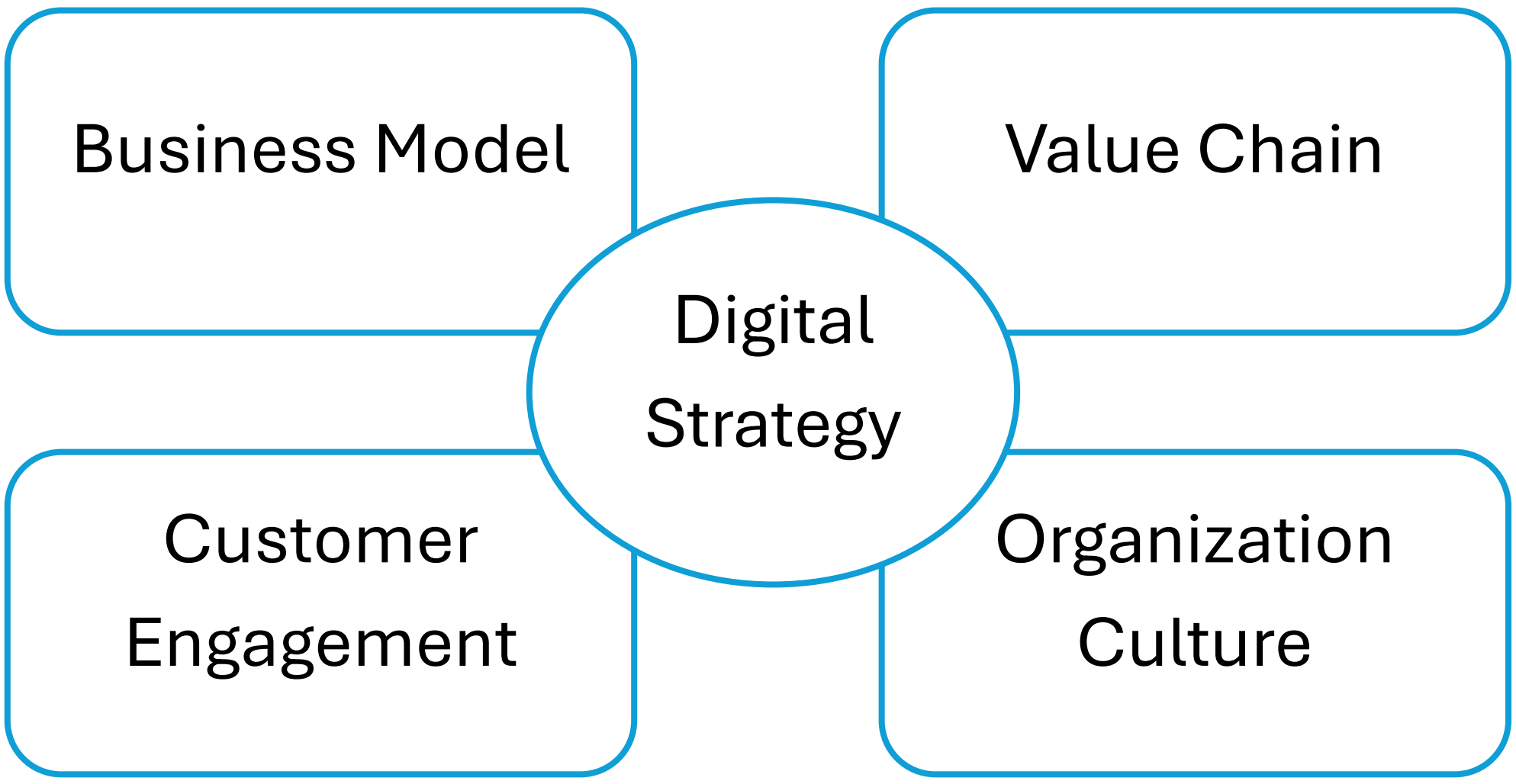

*Note.* Created by the author based on Gupta's (2018) four domains of digital transformation.

Business model transformation involves a fundamental redefinition of how a firm creates and captures value through digital integration. This encompasses the transition to new revenue streams, digital service extensions, platform participation, and data-driven value propositions (Bharadwaj et al., 2013; Gupta, 2018; Verhoef et al., 2021). While SMEs are frequently hampered by resource scarcity, digital technologies serve as a critical equalizer, enabling new forms of scalability and global customer reach that were previously inaccessible(Li et al., 2018; Nambisan et al., 2019). Ultimately, successful business model innovation in SMEs relies on the firm's ability to leverage these tools to drive operational efficiency while simultaneously exploring new market opportunities (Karadağ et al., 2024; Latifi et al., 2021).

Value chain transformation refers to the end-to-end digitalization and integration of internal and external processes, spanning procurement, production, logistics, and partner coordination. Digital technologies, specifically cloud systems, industrial automation, and real-time analytics, streamline workflows and enhance responsiveness by reducing informational silos (Gupta, 2018; Ivanov & Dolgui, 2021; Porter & Heppelmann, 2014)  For SMEs, digitalizing the value chain is a prerequisite for transparency and agility, enabling them to collaborate more effectively within larger digital ecosystems and respond dynamically to supply chain disruptions (Matarazzo et al., 2021; Rahman et al., 2025).

Customer Relationship Transformation involves leveraging digital tools to deepen engagement, automate Personalisation, and refine service delivery. By integrating digital channels, CRM systems, and data analytics, SMEs can move beyond intuition to a data-informed understanding of customer needs, allowing for highly tailored offerings (Gupta, 2018; Lemon & Verhoef, 2016). This domain is particularly vital for SMEs, as it allows them to scale their traditional strengths of relational proximity and personalized service, converting these informal advantages into sustainable, digitally enabled competitive assets (Ciasullo et al., 2025; Matarazzo et al., 2021).

Cultural transformation encompasses shifts in organisational norms, behaviours, and mindsets that support digital ways of working. Research highlights cultural attributes, such as experimentation, openness, collaboration, and data-driven decision-making, as critical enablers of digital transformation (Gupta, 2018; Warner & Wäger, 2019; Westerman et al., 2011). For SMEs, this transformation presents a unique paradox: while smaller firms may resist change due to entrenched routines, they also possess the structural flexibility to embed new cultural norms more rapidly than larger, more bureaucratic organizations (Karadağ et al., 2024; Khin & Ho, 2019).

Taken together, these four domains, business model, value chain, customer relationships, and culture, represent the structural and behavioural foundations of the organisation. Transformation across these domains is essential for SMEs to move beyond surface-level digital adoption toward deeper, more integrated forms of renewal. In the context of this paper, organisational domain transformation is conceptualised as the central organisational layer linking Digital Transformation Practices (DASAT) with the adoption of ExO-like attributes. Through changes in these domains, SMEs may create the organisational conditions that support greater adaptability, responsiveness, and non-linear growth potential.

### 2.4 ExO-like Attribute Adoption

The Exponential Organizations (ExO) framework identifies eleven attributes, Massive Transformative Purpose (MTP), Staff on Demand, Community & Crowd, Algorithms, Leveraged Assets, Engagement, Interfaces, Dashboards, Experimentation, Autonomy, and Social Technologies, that enable digitally native, platform-based firms to scale rapidly through external resources, algorithmic optimisation, and community-driven engagement (Ismail et al., 2014, 2023). However, most of these attributes, such as Staff on Demand, Leveraged Assets, Interfaces, Algorithms, and Social Technologies, presuppose platform architectures, large user bases, or venture-scale resources that are typically unavailable to SMEs. Recent scholarship emphasises that while SMEs cannot adopt the full ExO model, they can selectively implement a subset of ExO-inspired mechanisms that enhance agility, learning, and responsiveness (Bresciani et al., 2021; Khin & Ho, 2019; Scarmozzino et al., 2017).

Figure #3 provides a visual overview of the Exponential Organizations (ExO) framework, depicting the eleven attributes identified by Ismail et al. (2014, 2023)  that underpin scalable, platform-based growth.

**Figure #3**

*The Exponential Organization (ExO) Framework and Its Eleven Attributes*

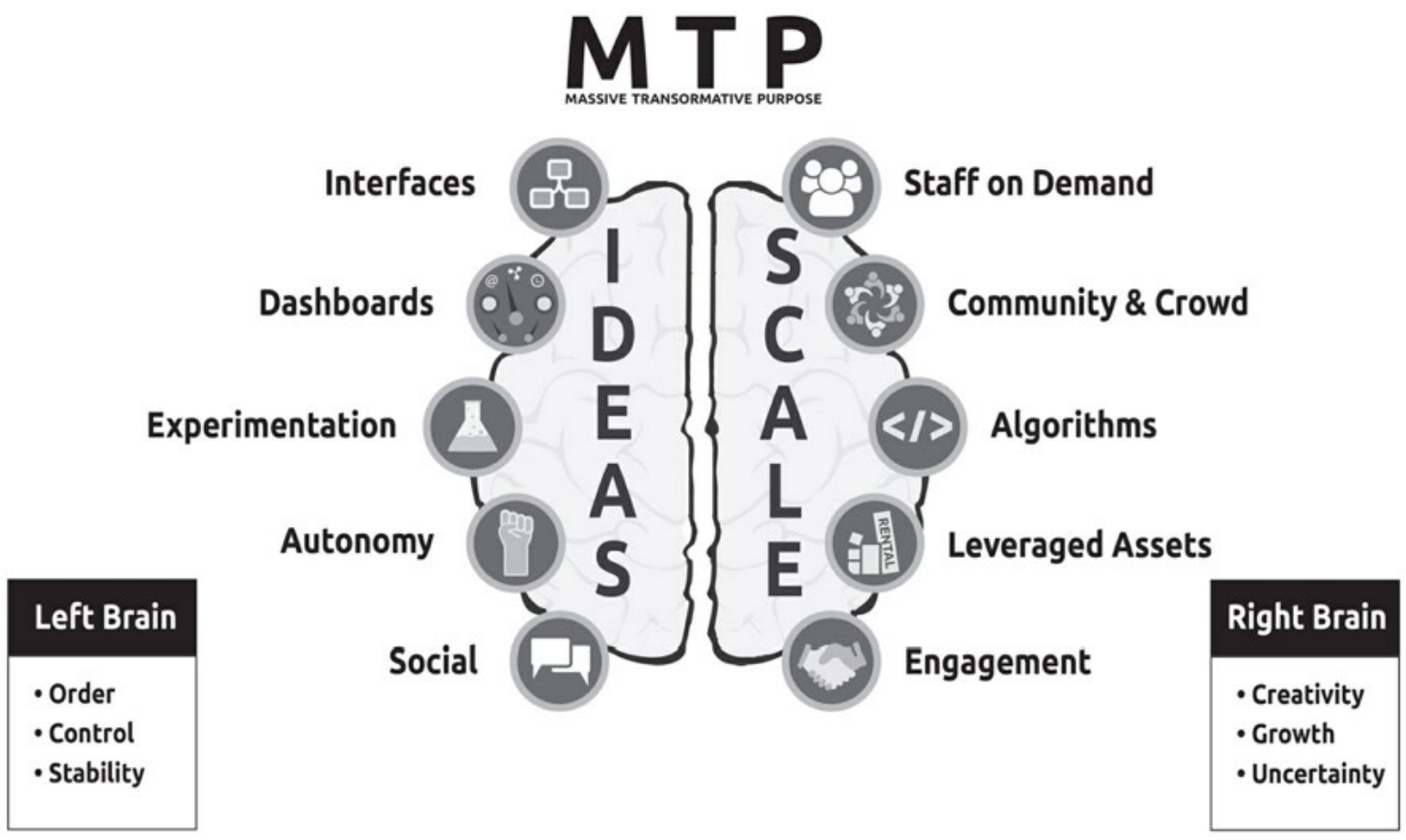

*Note.* Adapted from *Exponential Organizations: Why New Organizations Are Ten Times Better, Faster, and Cheaper Than Yours (and What to Do About It)* (Ismail et al., 2014). Copyright 2014 by Diversion Books. Adapted for academic discussion purposes.

Consistent with this empirical reality, this paper focuses on a subset of ExO-like attributes that meet three criteria. First, they are feasible for SMEs to adopt given typical resource constraints. Prior research shows that SMEs rely on lean, incremental, and low-cost mechanisms rather than data-intensive or algorithmic architectures (Matarazzo et al., 2021; Scarmozzino et al., 2017). Second, they are empirically supported as mechanisms that enhance adaptability and performance in smaller firms. Experimentation has been shown to accelerate learning and innovation (Bresciani et al., 2021; Thomke, 2020), while dashboards and data-driven decision-making improve responsiveness and operational clarity (Brynjolfsson & McElheran, 2016). Furthermore, autonomy increases organisational agility (Lee & Edmondson, 2017), and community engagement enables SMEs to access external knowledge and resources(Nambisan

& Baron, 2021). Third, these attributes align with the conceptually ordered relationships proposed within the Exponential SME Framework. Advanced digital mechanisms typically emerge only after foundational changes in business models, value chains, customer relationships, and culture (Verhoef et al., 2021; Warner & Wäger, 2019). These four attributes therefore represent mechanisms that SMEs can realistically adopt once domain-level transformation has occurred.

Based on these criteria, four ExO-like attributes are selected as theoretically appropriate and practically attainable for SMEs: experimentation, dashboards, autonomy, and community engagement. The selection of these attributes should not be interpreted as a rejection of the broader ExO framework. Rather, it reflects the contextual realities of SMEs, where organisational scale, resource constraints, and digital maturity often limit the applicability of several ExO attributes originally developed for digitally native platform organisations (Bresciani et al., 2021; Ismail et al., 2023). Consequently, the attributes selected in this study represent those most theoretically relevant and practically attainable within the SME context.

These attributes do not require platform scale or algorithmic infrastructure, yet they meaningfully enhance organisational adaptability and learning. Within the Exponential SME Framework, they represent organisational mechanisms through which transformed organisational domains may contribute to improved agility, responsiveness, and ultimately non-linear growth potential.

As discussed earlier, the remaining ExO attributes, such as Staff on Demand, Leveraged Assets, Algorithms, Interfaces, and Social Technologies, presuppose platform architectures, large user bases, or advanced data capabilities that are typically absent in SMEs. Their exclusion is therefore consistent with both theoretical boundary conditions and empirical evidence. The four selected attributes provide a focused, contextually grounded set of mechanisms that align with the SME transformation pathway developed in this paper and serve as the final stage in the proposed conceptual model.

Experimentation involves rapid testing, iterative learning, and low-cost trials to explore new ideas while reducing market uncertainty (Thomke, 2020). While SMEs often lack formal, capital-intensive R&D departments, experimentation serves as a "dynamic capability" that enables them to innovate incrementally, validate assumptions, and pivot with minimal sunk costs (Bresciani et al., 2021; Khin & Ho, 2019). However, effective experimentation is not merely technical; it requires cultural norms that tolerate risk and digitalized processes that facilitate rapid, data-informed iterations (Rahman et al., 2025).

Dashboards and data-driven decision-making involve the use of real-time metrics and analytics to guide both operational and strategic choices. By providing a centralized view of performance, digital dashboards enhance transparency and coordination, enabling SMEs to overcome the delays inherent in manual reporting (Brynjolfsson & McElheran, 2016; Karadağ et al., 2024). However, these tools only become meaningful strategic assets once an SME has successfully digitalized its core processes and fostered an organisational culture that prioritizes evidence over intuition(Rahman et al., 2025). Without this foundational digital maturity, dashboards risk becoming "vanity metrics" rather than drivers of agility.

Autonomy refers to decentralized decision-making and the empowerment of teams to act without constant top-down intervention. By reducing bureaucratic delays, autonomy enhances organisational agility, enabling employees to respond rapidly to emerging market opportunities and localized challenges (Lee & Edmondson, 2017; Warner & Wäger, 2019). While SMEs inherently possess flatter structures that facilitate the implementation of autonomous work-patterns, successful adoption still requires a deep cultural transformation and absolute clarity of strategic intent to ensure decentralized actions align with firm-level goals (Khin & Ho, 2019).

Community engagement involves leveraging digital platforms, social networks, and external ecosystems to co-create value, access distributed knowledge, and extend organisational reach (Nambisan & Baron, 2021). By engaging with these networks, SMEs can strengthen customer relationships and access specialized resources that exist beyond their internal boundaries, effectively compensating for their smaller scale (Matarazzo et al., 2021). However, community engagement acts as a force multiplier only when the firm's business model and customer relationship domains have already undergone digital transformation, providing the necessary infrastructure to capture and process external value.

These four ExO-like attributes are conceptualised as downstream mechanisms that SMEs can adopt only after sufficient transformation across key organisational domains. They do not emerge spontaneously; rather, they require prior changes in business models, value chains, customer relationships, and organisational culture. In this model, ExO-like attribute adoption represents the point at which SMEs begin to leverage digital transformation not only for efficiency but also for adaptability, innovation, and strategic renewal. Once adopted, these mechanisms enhance organisational agility and responsiveness, thereby contributing to the firm’s potential for non-linear growth.

## 2.5 Non-linear Growth Potential

Non-linear growth refers to disproportionate improvements in organisational performance that exceed what would typically be expected through incremental resource accumulation or gradual operational improvements (McGrath, 2013; Sirmon et al., 2011). Such growth is often characterised by step-change outcomes arising from innovation, strategic renewal, organisational reconfiguration, or the creation of new value-generating mechanisms. Unlike linear growth, which generally reflects proportional increases in inputs and outputs, non-linear growth enables firms to achieve substantially greater performance gains relative to their existing resource base. For SMEs, non-linear growth does not necessarily imply the exponential scaling associated with digitally native platform firms. Rather, it reflects the capacity to achieve disproportionate value creation, enhanced competitiveness, and accelerated performance outcomes through digitally enabled organisational transformation (Karadağ et al., 2024; Nambisan et al., 2017). Figure #4 illustrates the distinction between linear growth trajectories and the more accelerated performance patterns associated with non-linear growth.

**Figure #4**

*Linear and Non-linear Growth Pattern*

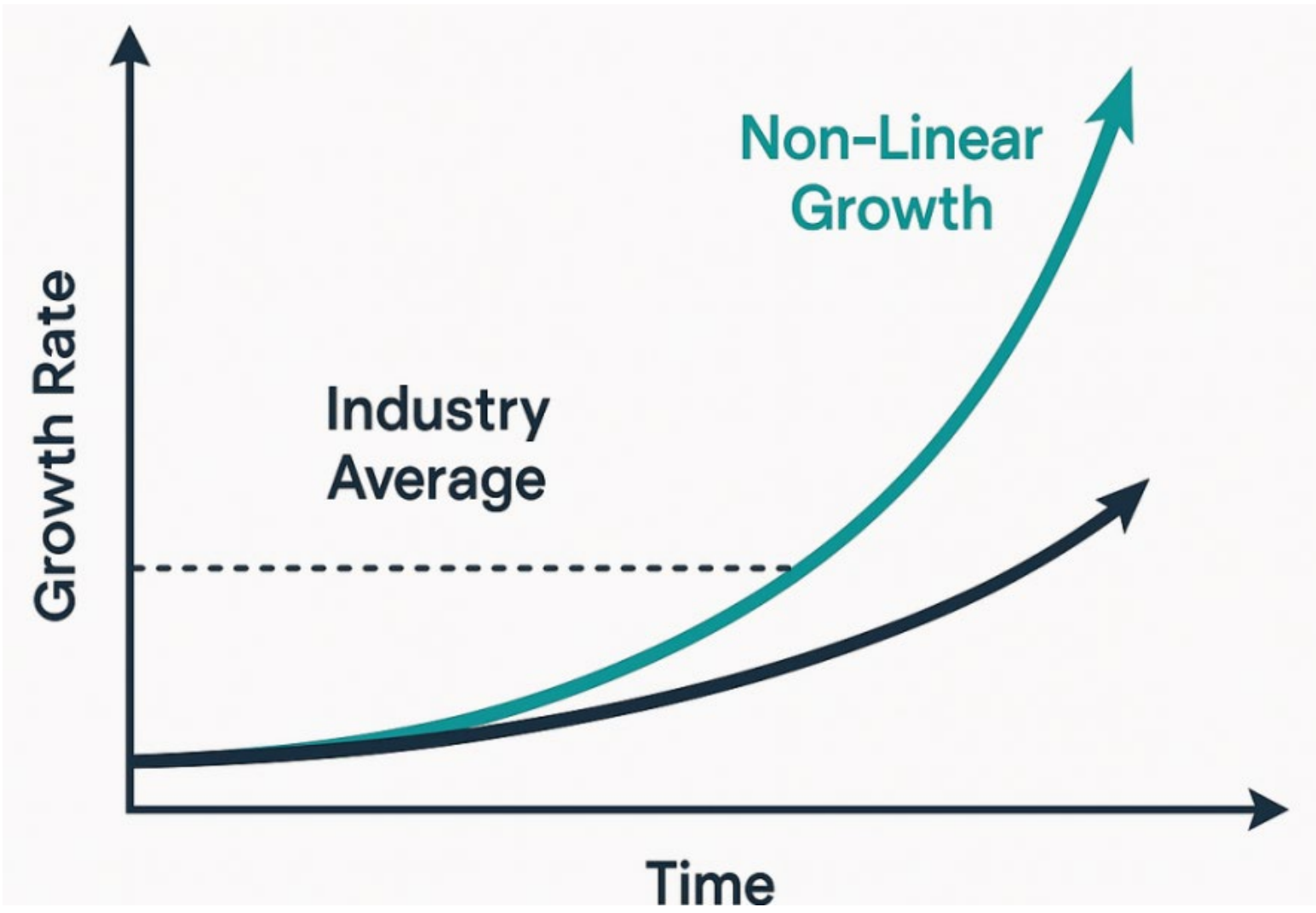

*Note.* Created by the author

Recent SME growth research suggests that growth trajectories are often heterogeneous and episodic, with firms experiencing periods of acceleration, consolidation, renewal, and temporary stagnation rather than predictable linear progression. Research on strategic agility, dynamic capabilities, and resource orchestration similarly suggests that organisational growth is shaped by firms' ability to adapt, reconfigure resources, and respond to changing environments rather than simply accumulate resources over time (Doz & Kosonen, 2010; Sirmon et al., 2011; D. J. Teece, 2007) . These observations reinforce the need to understand how organisational capabilities, transformation practices, and structural adaptations interact to support growth-enhancing transformation pathways.

Digital technologies play a central role in enabling non-linear growth by reducing transaction costs, expanding market access, and enhancing organisational agility (Bharadwaj et al., 2013). However, technology alone is insufficient; SMEs require complementary organisational changes, specifically across business models, digitalised value chains, customer engagement strategies, and adaptive cultural norms, to translate digital investments into meaningful growth outcomes (Gupta, 2018; Verhoef et al., 2021). These domain-level transformations create the structural and behavioural conditions that may enable SMEs to leverage digital tools in ways that support step-change improvements in performance rather than marginal gains.

Non-linear growth potential is fundamentally shaped by a firm's ability to innovate, experiment, and adapt rapidly to shifting market environments. Research on organisational agility and dynamic capabilities highlights that firms capable of sensing emerging opportunities, seizing them through coordinated strategic action, and reconfiguring resources can achieve disproportionate competitive advantages (Doz & Kosonen, 2010; D. Teece et al., 2016). For SMEs, this suggests that growth is less a function of sheer physical scale and more a function of strategic responsiveness, learning capacity, and digitally enabled adaptability.

In this paper, non-linear growth potential is conceptualised as the ultimate outcome of the transformation pathway. It represents the firm's enhanced ability to generate step-change improvements in performance through the adoption of ExO-like mechanisms, such as experimentation, dashboards, autonomy, and community engagement, made possible by prior transformation of key organisational domains. These mechanisms amplify the effects of digital transformation by enabling SMEs to innovate more rapidly, make better decisions, and mobilise internal and external resources more effectively.

Importantly, non-linear growth potential is not guaranteed; it is contingent on the alignment of digital capabilities, transformation practices, domain-level transformations, and ExO-like attribute adoption. Within the Exponential SME Framework, non-linear growth is viewed as an emergent organisational outcome arising from the interaction of multiple complementary organisational mechanisms rather than any single capability, technology, or intervention.

## 3. The Exponential SME Framework

The preceding sections synthesised literature relating to Digital Dynamic Capabilities (DDC), Digital Transformation Practices (DASAT), Organisational Domain Transformation, ExO-like Attributes, and Non-linear Growth Potential. While each of these streams has generated valuable insights, they have largely evolved independently. Consequently, there remains limited understanding of how these concepts collectively contribute to digitally enabled growth within SMEs.

To address this gap, this paper proposes the Exponential SME Framework, an integrated conceptual framework that brings together these previously fragmented streams of literature. The framework provides a holistic view of how SMEs may leverage digital capabilities, enact transformation practices, reshape key organisational domains, adopt growth-enabling organisational mechanisms, and enhance their potential for non-linear growth.

Rather than representing a deterministic or universally applicable sequence, the framework synthesises relationships identified within the literature and organises them into a conceptually ordered structure that reflects how digital transformation is commonly understood to unfold within SMEs. The framework is intended as a synthesis of existing knowledge rather than a causal theory of SME digital transformation and growth.

### 3.1 Digital Dynamic Capabilities and Digital Transformation Practices (DASAT)

Digital Dynamic Capabilities (DDC) represent a firm's capacity to sense digital opportunities, seize digital initiatives, and reconfigure organisational resources in response to technological and environmental change (Teece, 2007; Warner & Wäger, 2019). Within SMEs, these capabilities provide the organisational foundation required to recognise digital opportunities and mobilise change despite resource constraints (Ciasullo et al., 2025; Karadağ et al., 2024) .

However, digital capabilities alone do not constitute transformation. Capabilities must be enacted through organisational practices that translate digital intent into action. The DASAT framework, comprising Digital Awareness, Digital Strategy, Digital Adoption, and Continuous Improvement, captures these practices and provides a practical representation of how digital transformation is implemented within SMEs.

Drawing on the literature, DDC can therefore be viewed as the foundational capabilities that support the enactment of DASAT. Firms with stronger digital capabilities are more likely to engage in systematic environmental scanning, strategic alignment, digital adoption, and continuous organisational learning (Chanias et al., 2019; Kahveci, 2025; Matarazzo et al., 2021).

**Relationship 1:** Digital Dynamic Capabilities provide the foundation for the enactment of Digital Transformation Practices (DASAT) within SMEs.

### 3.2 Digital Transformation Practices and Organisational Domain Transformation

While DASAT captures the practices through which digital transformation is enacted, sustainable organisational change requires transformation across core organisational domains.

These domains include business models, value chains, customer relationships, and organisational culture (Gupta, 2018; Verhoef et al., 2021).

The literature suggests that the four DASAT practices contribute to organisational transformation in different but complementary ways. Digital Awareness enables firms to recognise emerging opportunities and changing customer expectations. Digital Strategy provides direction and alignment for transformation initiatives. Digital Adoption embeds new technologies and processes into organisational routines. Continuous Improvement supports ongoing organisational adaptation and learning.

Collectively, these practices may contribute to broader transformations that reshape how SMEs create, deliver, and capture value (Chanias et al., 2019; Verhoef et al., 2021; Vial, 2019) Through this lens, DASAT serves as the organisational mechanism through which digital capabilities are expressed and translated into more substantive organisational change.

**Relationship 2:** Digital Transformation Practices (DASAT) contribute to Organisational Domain Transformation within SMEs.

### 3.3 Organisational Domain Transformation and ExO-like Attribute Adoption

The ExO literature identifies several organisational attributes associated with adaptability, innovation, and responsiveness, including experimentation, dashboards, autonomy, and community engagement (Ismail et al., 2014, 2023). However, the effective adoption of these attributes requires supporting organisational structures, processes, and cultural conditions.

The transformation of business models, value chains, customer relationships, and organisational culture may create these enabling conditions. For example, business model transformation may encourage experimentation and innovation. Digitalised value chains may support real-time dashboards and increased transparency. Customer relationship transformation may strengthen community engagement. Cultural transformation may create environments supportive of autonomy and experimentation.

Within the Exponential SME Framework, organisational domain transformation therefore provides the context within which ExO-like attributes become both feasible and effective (Bresciani et al., 2021; Warner & Wäger, 2019).

**Relationship 3:** Organisational Domain Transformation supports the adoption of ExO-like attributes within SMEs.

### 3.4 ExO-like Attribute Adoption to Non-linear Growth Potential

The ExO-like attributes identified in this framework, experimentation, dashboards, autonomy, and community engagement, are associated with enhanced organisational adaptability, learning, and responsiveness (Bresciani et al., 2021; Ismail et al., 2014, 2023).

Experimentation supports rapid learning and innovation. Dashboards improve decision quality through greater information visibility. Autonomy enables faster responses to opportunities and challenges. Community engagement expands access to external knowledge, resources, and networks.

Within the Exponential SME Framework, these attributes are not viewed as direct drivers of growth. Rather, they represent organisational mechanisms that may strengthen the conditions associated with non-linear growth. Through enhanced adaptability, innovation, and responsiveness, SMEs may improve their capacity to generate disproportionate performance gains relative to their resource base.

**Relationship 4:** The adoption of ExO-like attributes strengthens the organisational conditions associated with non-linear growth potential within SMEs.

### 3.5 The Exponential SME Framework

The framework proposes that Digital Dynamic Capabilities provide the foundation for Digital Transformation Practices, which contribute to Organisational Domain Transformation. These transformed domains may support the adoption of ExO-like attributes that strengthen the conditions associated with non-linear growth potential. Importantly, the framework does not suggest that growth outcomes are determined by any single capability, practice, or intervention. Rather, non-linear growth potential is viewed as an emergent outcome arising from the interaction of multiple complementary organisational mechanisms. Collectively, the framework positions non-linear growth potential as an emergent outcome arising from the interaction of capabilities, transformation practices, organisational transformation, and growth-enabling organisational mechanisms rather than any single factor in isolation.

Figure #5 presents the Exponential SME Framework and provides a visual representation of the relationships synthesised throughout this paper.

**Figure #5**

*The Exponential SME Framework*

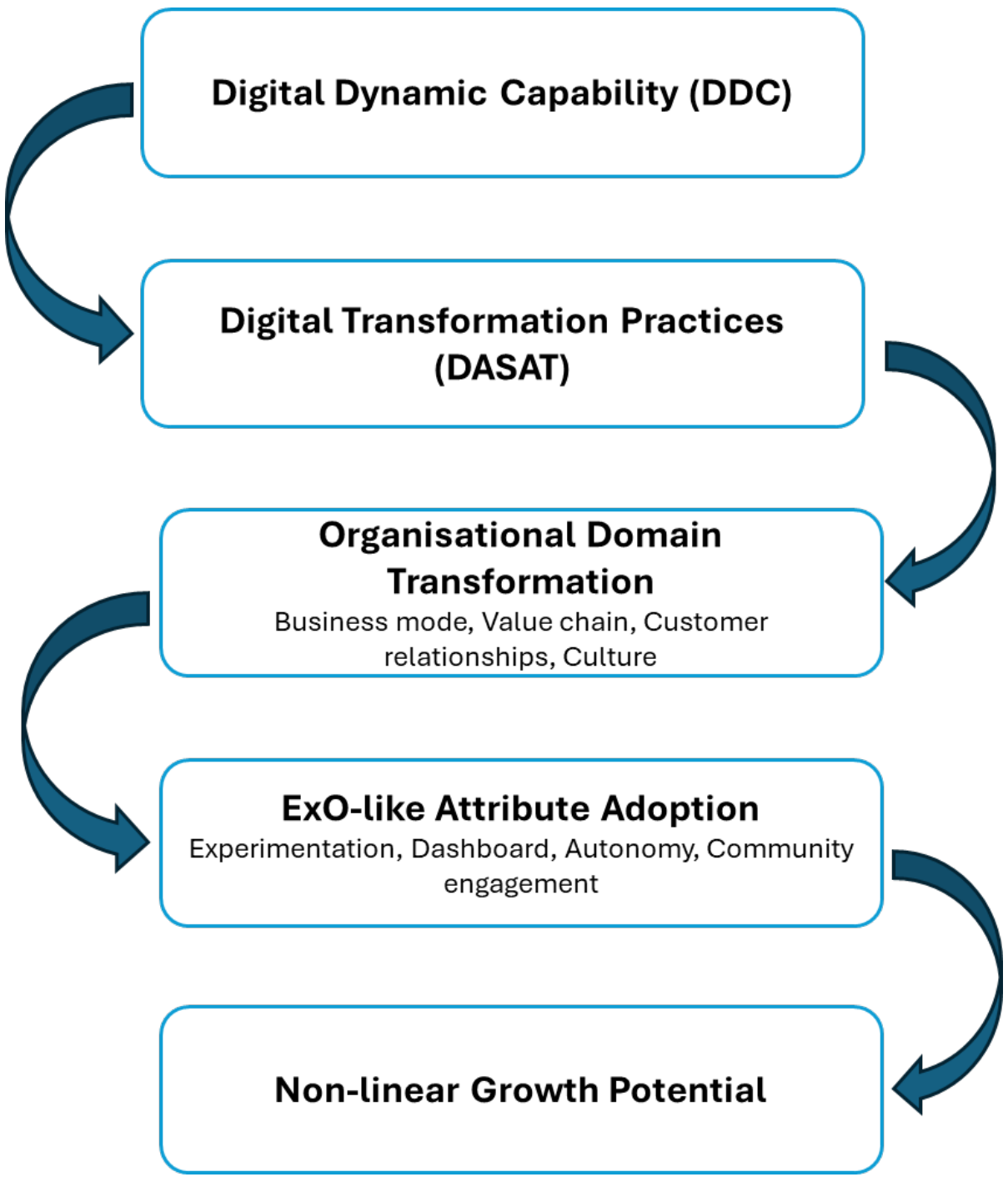

*Note.* Created by the author. The Exponential SME Framework synthesises literature relating to Digital Dynamic Capabilities (Teece, 2007), Digital Transformation Practices (Kahveci, 2025), Organisational Domain Transformation (Gupta, 2018), and ExO-like Attributes (Ismail et al., 2014, 2023).

## 4. Discussion

This paper set out to address a persistent gap within the SME digital transformation literature: the absence of an integrated explanation of how digital capabilities, transformation practices, organisational transformation, and growth-enabling mechanisms collectively contribute to non-linear growth potential. While substantial research has examined these domains independently, relatively little attention has been devoted to understanding how they may interact within a broader SME transformation context. In response, this paper developed the Exponential SME Framework, a conceptual framework that synthesises literature on Digital Dynamic Capabilities (DDC), Digital Transformation Practices (DASAT), Organisational Domain Transformation, ExO-like Attributes, and Non-linear Growth Potential.

The framework contributes to the literature by consolidating fragmented streams of research into a coherent structure that explains how SMEs may progress from digital capability development towards growth-enhancing organisational transformation. Rather than presenting a new theory of digital transformation, the framework provides an integrative lens through which existing knowledge can be understood collectively. In doing so, it offers a foundation for future empirical investigation and theory development within SME digital transformation research.

### 4.1 Integrating Fragmented Literature Streams

The primary contribution of the Exponential SME Framework lies in its integration of several research streams that have traditionally evolved independently. Research on Digital Dynamic Capabilities has focused primarily on how organisations develop the capacity to respond to technological change (Teece, 2007; Warner & Wäger, 2019) . Studies of digital transformation have examined the practices and activities through which transformation is enacted (Chanias et al., 2019; Vial, 2019) . Organisational transformation research has concentrated on changes to business models, value chains, customer relationships, and culture (Gupta, 2018; Verhoef et al., 2021). Meanwhile, ExO research has explored organisational attributes associated with adaptability, innovation, and scalability (Ismail et al., 2014, 2023).

While these streams have generated important insights, they have typically been examined in isolation. The Exponential SME Framework contributes by bringing these perspectives together and highlighting their potential relationships within a single conceptual structure. This integration advances understanding of SME digital transformation by demonstrating that growth-enhancing transformation cannot be adequately explained through capabilities, practices, organisational change, or ExO-like attributes alone. Rather, these elements appear to operate as interconnected components of a broader organisational transformation process.

### 4.2 Clarifying Relationships Within SME Digital Transformation

A second contribution of the framework is its articulation of the relationships linking digital capabilities, transformation practices, organisational transformation, and growth-enabling organisational mechanisms. Existing research has often examined these elements separately, making it difficult to understand how they interact within SMEs.

The Exponential SME Framework suggests that Digital Dynamic Capabilities provide a foundation upon which transformation practices are enacted. These practices may contribute to changes in core organisational domains, including business models, value chains, customer relationships, and culture. In turn, transformed organisational domains may create conditions that support the adoption of ExO-like attributes such as experimentation, dashboards, autonomy, and community engagement.

Importantly, the framework does not imply that all SMEs will experience transformation in a uniform or linear manner. Rather, it proposes a conceptually ordered structure that helps explain how different transformation elements may reinforce one another. This perspective contributes to ongoing discussions regarding organisational readiness, capability development, and digital transformation by highlighting the interdependence of multiple organisational layers.

### 4.3 Reframing SME Digital Transformation Through an Organisational Lens

The framework also contributes by shifting attention away from technology-centric explanations of SME digital transformation. Much of the digital transformation literature focuses on technology adoption as the primary driver of organisational change. In contrast, the Exponential SME Framework emphasises the organisational conditions that support transformation.

This perspective highlights the importance of capabilities, practices, organisational structures, cultural change, and growth-enabling mechanisms as complementary dimensions of transformation. By adopting this broader organisational lens, the framework provides a more holistic explanation of why some SMEs appear capable of leveraging digital transformation effectively while others struggle to generate meaningful outcomes despite similar technology investments.

The framework therefore aligns with emerging calls for more integrative and organisationally grounded approaches to digital transformation research. It reinforces the view that transformation is not simply a technological phenomenon but a multifaceted organisational process involving learning, adaptation, strategic alignment, and organisational renewal.

### 4.4 Implications for Understanding SME Growth

The framework also contributes to SME growth research by providing a conceptual explanation for how digital transformation may strengthen the conditions associated with non-linear growth potential. Traditional growth models have frequently focused on resource accumulation, market expansion, or strategic positioning. The Exponential SME Framework complements these perspectives by emphasising organisational adaptability, experimentation, learning, responsiveness, and ecosystem engagement.

Within the framework, non-linear growth is not viewed as a direct outcome of technology adoption. Instead, it is conceptualised as a potential organisational outcome that may emerge when capabilities, transformation practices, organisational transformation, and ExO-like attributes reinforce one another. This interpretation provides a useful lens for understanding how SMEs may achieve disproportionate performance improvements without necessarily possessing the scale and resources associated with larger organisations.

The framework therefore extends existing discussions of SME growth by linking growth potential to organisational transformation processes rather than solely to resource accumulation or market positioning.

### 4.5 Implications for Research, Practice, and Policy

The Exponential SME Framework has implications for future research, managerial practice, and policy development.

For researchers, the framework provides an integrative foundation for examining the relationships between digital capabilities, digital transformation practices, organisational transformation, ExO-like attributes, and growth outcomes. Future empirical studies can investigate these relationships, examine boundary conditions, and explore how transformation pathways differ across industries, organisational contexts, and levels of digital maturity.

For practitioners, the framework highlights that digital transformation should not be viewed as a series of isolated technology projects. Instead, transformation requires coordinated attention to capability development, transformation practices, organisational change, and growth-enabling mechanisms. This perspective may help SME leaders identify capability gaps, prioritise transformation initiatives, and evaluate organisational readiness.

For policymakers, the framework suggests that SME digitalisation programs should extend beyond technology adoption incentives. Effective support may also require interventions that strengthen managerial capabilities, organisational transformation capacity, experimentation, learning, and ecosystem participation. Such initiatives may improve the ability of SMEs to convert digital transformation efforts into sustainable growth outcomes.

# 5. Managerial and Policy Implications

The Exponential SME Framework offers several actionable insights for SME leaders and policymakers seeking to strengthen digital transformation outcomes and enhance non-linear growth potential. By highlighting the relationships between digital capabilities, transformation practices, organisational transformation, and growth-enabling organisational mechanisms, the framework identifies areas where interventions, investments, and support programs may have the greatest impact.

## 5.1 Managerial Implications

### 5.1.1 Strengthening Digital Dynamic Capabilities as a Strategic Priority

SME leaders should prioritise the development of Digital Dynamic Capabilities (DDC) as the foundation for broader digital transformation efforts. This includes investing in leadership digital literacy, establishing routines for environmental scanning, and building the strategic capacity to evaluate and prioritise digital opportunities. Without these foundational capabilities, digital initiatives risk becoming fragmented or technology-driven rather than strategically aligned.

### 5.1.2 Institutionalising Structured Digital Transformation Practices (DASAT)

Managers should adopt structured digital transformation practices, digital awareness, strategy, adoption, and continuous improvement, as repeatable organisational routines rather than one-off initiatives. This requires formalising digital strategy processes, establishing cross-functional transformation teams, and embedding iterative learning cycles. Institutionalising DASAT ensures that digital transformation becomes an ongoing organisational capability rather than a reactive response to external pressures.

### 5.1.3 Focusing on Domain-Level Transformation, Not Just Technology Adoption

The model emphasises that meaningful transformation requires changes across business models, value chains, customer relationships, and organisational culture. SME leaders should therefore evaluate digital initiatives based on their potential to reshape these domains, rather than on their technological sophistication alone. This may involve redesigning value propositions, digitising core processes, or cultivating a culture that supports experimentation and data-driven decision-making.

### 5.1.4 Creating the Conditions for ExO-like Attributes to Flourish

Experimentation, dashboards, autonomy, and community engagement cannot be implemented effectively without the right organisational foundations. Managers should ensure that data infrastructure, governance structures, and cultural norms are aligned with these advanced practices. For example, decentralised autonomy requires clear decision rights and trust-based leadership, while experimentation requires psychological safety and tolerance for failure.

### 5.1.5 Leveraging External Communities for Scalable Growth

Community engagement offers SMEs access to knowledge, resources, and markets that would otherwise be unattainable. Managers should actively cultivate partnerships, user communities, and ecosystem relationships that enable co-creation and shared value. This approach allows SMEs to scale impact without proportionally increasing internal resources, strengthening the conditions associated with non-linear growth.

## 5.2 Policy Implications

### 5.2.1 Building SME Digital Capability Foundations

Policymakers should design programs that strengthen SMEs' Digital Dynamic Capabilities, including leadership development, digital literacy training, and strategic capability building.

These foundational capabilities are prerequisites for effective digital transformation and should be prioritised over technology-subsidy programs that lack strategic alignment.

#### 5.2.2 Supporting Structured Transformation Practices

Government initiatives should encourage SMEs to adopt structured digital transformation practices by providing frameworks, diagnostic tools, and advisory services. Programs that help SMEs develop digital strategies, conduct capability assessments, or implement continuous improvement routines can significantly accelerate transformation readiness.

#### 5.2.3 Investing in Digital Infrastructure and Data Accessibility

The adoption of ExO-like attributes, particularly dashboards and autonomy, depends on reliable digital infrastructure and accessible data. Policymakers should therefore invest in regional digital infrastructure, promote interoperability standards, and support SMEs in adopting data governance practices that enable real-time decision-making.

#### 5.2.4 Encouraging Ecosystem-Based Innovation and Community Engagement

Policies that foster collaboration between SMEs, universities, industry associations, and innovation hubs can strengthen community engagement and knowledge sharing. Incentives for open innovation, platform participation, and ecosystem partnerships can help SMEs access external resources and accelerate non-linear growth trajectories.

#### 5.2.5 Designing Stage-Appropriate Support Programs

Because SMEs differ considerably in their digital capabilities, transformation practices, and organisational readiness, policy interventions should be tailored to the firm's level of digital transformation maturity. Early-stage firms may require capability development and strategic guidance, while more advanced firms may benefit from support for experimentation, data analytics, or ecosystem integration. Stage-appropriate interventions increase the likelihood of successful transformation and reduce wasted resources.

## 6. Future Research Directions

The Exponential SME Framework opens several promising avenues for future research on SME digital transformation, capability development, organisational transformation, and non-linear growth. By integrating previously fragmented streams of literature, the framework highlights conceptual gaps and empirical opportunities that can advance understanding of how SMEs evolve in increasingly digital environments.

### 6.1 Empirical Examination of the Exponential SME Framework

A primary opportunity lies in empirically examining the relationships articulated within the Exponential SME Framework. While each relationship is grounded in established literature, the integrated framework linking Digital Dynamic Capabilities (DDC), Digital Transformation Practices (DASAT), Organisational Domain Transformation, ExO-like Attribute Adoption, and Non-linear Growth Potential has not yet been examined holistically.

Future studies could employ:

- Longitudinal designs to capture capability evolution over time
- Structural equation modelling to examine relationships across the framework
- Cross-lagged panel analyses to explore temporal dynamics
- Mixed-methods approaches to investigate the underlying organisational processes and interactions

Such studies would provide empirical clarity regarding the nature, strength, and contextual variation of the relationships proposed within the framework.

### 6.2 Exploring Boundary Conditions and Moderators

The framework presents a generalised representation of SME transformation. However, contextual factors may shape how the proposed relationships unfold in practice. Future research could examine moderating influences such as:

- Industry turbulence (e.g., fast-moving versus stable sectors)
- Resource constraints (financial, human, and technological)
- Leadership digital orientation
- Firm age and growth stage
- Ecosystem embeddedness

Understanding these contextual influences would help refine the framework and identify circumstances under which SME transformation may accelerate, stall, or follow alternative trajectories.

### 6.3 Micro-Foundations of ExO-like Attribute Adoption

While the framework identifies four ExO-like attributes relevant to SMEs, the organisational micro-foundations that support their adoption remain underexplored. Future research could investigate:

- How psychological safety and trust influence experimentation
- How governance structures and decision rights support autonomy
- How data literacy shapes dashboard utilisation and effectiveness
- How community engagement emerges and is sustained within SMEs

Qualitative studies, ethnographic approaches, and in-depth case research would be particularly valuable in uncovering these behavioural, cultural, and organisational dynamics.

### 6.4 Measuring Non-linear Growth Potential

Non-linear growth potential is conceptually rich but empirically challenging to operationalise and measure. Future research could develop and validate metrics capable of capturing:

- Disproportionate performance gains
- Scalability without proportional resource expansion
- Ecosystem-driven value creation
- Organisational agility and responsiveness under changing conditions

Developing robust measurement approaches would strengthen future empirical investigations of non-linear growth and contribute to greater conceptual clarity within the SME growth literature.

### 6.5 Comparative Studies Across Firm Sizes and Ecosystems

The Exponential SME Framework has been developed specifically for SMEs. However, comparative research could examine how the proposed relationships differ across organisational contexts, including:

- Micro-enterprises
- Mid-sized firms
- Large organisations
- Platform-based firms
- Firms embedded within innovation ecosystems

Such comparative studies could help identify which aspects of the framework are SME-specific and which may have broader applicability across different organisational settings.

### 6.6 Policy and Ecosystem-Level Research

Future research could also explore how policy interventions, ecosystem structures, and institutional environments influence SME transformation outcomes. Potential research questions include:

- How do investments in regional digital infrastructure influence organisational transformation?
- Which policy interventions are most effective in strengthening Digital Dynamic Capabilities within SMEs?
- How do innovation hubs, accelerators, universities, and industry associations support experimentation and community engagement?

Such research would deepen understanding of how external environments enable or constrain SME digital transformation and growth. It would also provide valuable insights for policymakers seeking to strengthen national and regional digital capability development among SMEs.

## 7. Conclusion

This paper developed the Exponential SME Framework, an integrated conceptual framework that explains how digital capabilities, transformation practices, organisational transformation, and ExO-like attributes may collectively contribute to non-linear growth potential within SMEs. By synthesising insights from Digital Dynamic Capabilities (DDC), digital transformation research, organisational transformation literature, Exponential Organization (ExO) research, and SME growth scholarship, the framework provides a holistic perspective on how SMEs may leverage digital transformation to strengthen organisational adaptability and growth readiness.

A key contribution of the framework lies in its integration of previously fragmented literature streams. Rather than focusing on capabilities, practices, organisational change, or growth outcomes in isolation, the Exponential SME Framework highlights the relationships between these elements and positions them within a coherent conceptual structure. In doing so, the framework advances understanding of SME digital transformation as a multi-layered organisational phenomenon involving the interaction of capabilities, practices, organisational transformation, and growth-enabling organisational mechanisms.

The framework further contributes by reframing digital transformation beyond technology adoption alone. It suggests that successful transformation involves the development of digital capabilities, the enactment of structured transformation practices, the transformation of core organisational domains, and the adoption of organisational attributes that support agility, learning, experimentation, and external engagement. Through this lens, non-linear growth potential is viewed not as a direct outcome of technology adoption but as an emergent organisational outcome arising from the interaction of multiple complementary organisational elements.

For practitioners and policymakers, the framework highlights the importance of investing in capability development, institutionalising structured transformation practices, supporting organisational transformation, and creating the conditions that enable experimentation, autonomy, data-informed decision-making, and ecosystem participation. These insights are particularly relevant for SMEs operating under resource constraints, where transformation requires careful alignment between strategic intent, organisational capabilities, and growth ambitions.

The framework also provides a foundation for future empirical research. Opportunities remain to examine the relationships proposed within the framework, explore contextual influences and boundary conditions, investigate the micro-foundations of ExO-like attribute adoption, and develop more robust measures of non-linear growth potential. Such research will enhance understanding of how SMEs navigate digital transformation and how digitally enabled organisational transformation contributes to longer-term growth outcomes.

Ultimately, the Exponential SME Framework offers a foundation for advancing research, informing practice, and supporting policy development in the fields of SME digital transformation and growth. By providing an integrated perspective on how digital capabilities, transformation practices, organisational transformation, and ExO-like attributes interact, the framework contributes to a more comprehensive understanding of how SMEs may strengthen their capacity for non-linear growth in increasingly digital environments.

## Declaration of generative AI and AI-assisted technologies in the manuscript preparation process

During the preparation of this work, the authors used Microsoft Copilot to assist with language refinement, grammar, clarity, and the organisation of manuscript content. The tool was used solely to improve the presentation and readability of the manuscript and was not used to generate the study's intellectual contribution, theoretical arguments, or conclusions. Following the use of this tool, the authors critically reviewed, edited, and verified all content and take full responsibility for the accuracy, integrity, and originality of the work presented in this article.